\documentclass[epjCONF]{svjour}
\usepackage{graphicx}
\usepackage[varg]{txfonts} 
\usepackage[latin1]{inputenc}
\session-title{%
19$^{\textnormal{\footnotesize th}}$ International %
IUPAP Conference on Few-Body Problems in Physics%
}
\begin{document}
\title{%
Electromagnetic properties of the Beryllium-11 nucleus in Halo EFT
}%
\author{%
D.~R.~Phillips\inst{1,2}\fnmsep\thanks{\email{phillips@phy.ohiou.edu}} 
\and %
H.-W.~Hammer\inst{2} 
}
\institute{%
Department of Physics and Astronomy, Ohio University, Athens, Ohio 45701, USA
\and
Helmholtz-Institut f\"ur Strahlen- und Kernphysik and Bethe Center for Theoretical Physics, Universit\"at Bonn, D-53115, Bonn, Germany
}
\abstract{We compute electromagnetic properties of the Beryllium-11 nucleus using an effective field theory that exploits the separation of scales in this halo system. We fix the parameters of the EFT from measured data on levels and scattering lengths in the ${}^{10}$Be plus neutron system. We then obtain predictions for the B(E1) strength of the $1/2^+$ to $1/2^-$ transition in the ${}^{11}$Be nucleus. We also compute the charge radius of the ground state of ${}^{11}$Be. Agreement with experiment within the expected accuracy of a leading-order computation in this EFT is obtained. We also indicate how higher-order corrections that affect both s-wave and p-wave ${}^{10}$Be-neutron interactions will affect our results.} 
\maketitle
%
%
%
\section{Introduction}
\label{intro}


The first excitation of the Beryllium-10 nucleus is 3.4 MeV above the ground state, and that ground state has quantum numbers $0^+$. Meanwhile, 
the Beryllium-11 nucleus has a $1/2^+$ state whose neutron separation energy is 504 keV, and a $1/2^-$ state whose neutron separation energy is 184 keV. The shallowness of these two states of ${}^{11}$Be compared to the bound states of ${}^{10}$Be suggests that they have significant components in which a loosely-bound neutron orbits a ${}^{10}$Be core. In this ``one-neutron halo" picture the $1/2^+$ is predominantly an s-wave bound state, while the $1/2^-$ is predominantly a relative p-wave between the neutron and the core. 
In this paper we discuss efforts to use effective field theory (EFT) to systematically implement such a halo picture of the ${}^{11}$Be nucleus. 
 
This halo viewpoint is reinforced by the fact that the scattering volume of n${}^{10}$Be scattering in the $l=1$, $j=1/2$ channel has been determined to be~\cite{TB04}
\begin{equation}
a_1=457 \pm 67~{\rm fm}^3.
\label{eq:pwavescattlength}
\end{equation}
The corresponding length scale of order 8 fm is large compared to the natural length-scale of core-neutron interactions, which is $\approx 2$--$3$ fm. 

The datum (\ref{eq:pwavescattlength}), together with the information on the bound-state energies in the ${}^{10}$Be and ${}^{11}$Be systems, helps us to estimate the expansion parameter in our Halo EFT. This is the binding energy of the halo nucleus, as compared to the energy required to excite the core, i.e. $B_{\rm lo}/B_{\rm hi} \approx 1/6$. Converting this to an estimate of the different distance scales involved, we infer that a majority of the probability density of ${}^{11}$Be occupies a region outside the ${}^{10}$Be core: $R_{\rm core}/R_{\rm halo}$ $\approx 0.4$, which is consistent with the ratio implied by the numbers in the previous paragraph. This ratio of distance scales is the formal expansion parameter for the EFT, and since it is not particularly small, the leading-order calculations presented here are only a first step. Eventually sub-leading orders must be computed in order  to verify that the series is converging in the expected manner. 

Here we will apply this EFT to electromagnetic reactions in the ${}^{11}$Be system. The B(E1)($1/2^+ \rightarrow 1/2^-$) transition has recently been measured to be 
\begin{equation}
{\rm B(E1)}(1/2^+ \rightarrow 1/2^-)=0.105(12)~e^2~{\rm fm}^2
\label{eq:CEXBE1}
\end{equation}
using intermediate-energy Coulomb excitation~\cite{Su07}. This is consistent with the older number 
\begin{equation}
{\rm B(E1)}(1/2^+ \rightarrow 1/2^-)=0.116(12)~e^2~{\rm fm}^2
\end{equation}
from lifetime measurements~\cite{Mi83}. There are also two recent data sets on the Coulomb-induced breakup of the ${}^{11}$Be nucleus~\cite{Pa03,Fu04} (see also Ref.~\cite{An94}). Both experiments extracted the excitation function $d$B(E1)$/dE$ as a function of the energy of the outgoing neutron, $E$. For low neutron energies this excitation function is affected by the final-state interaction in the p-waves, and can be predicted in the halo picture~\cite{TB04}. The non-energy-weighted sum rule, applied to the data of Ref.~\cite{Pa03}, then gives a measurement of the radius of the ground state of ${}^{11}$Be of
\begin{equation}
\langle r^2 \rangle^{1/2}=5.7(4)~{\rm fm}.
\end{equation}
This is consistent with the recent atomic-physics measurement of the charge radius~\cite{No09}:
\begin{equation}
r_c=2.463(16)~{\rm fm},
\end{equation}
when one accounts for the fact that the matter radius and the charge radius differ by a factor of the effective charge, $Z_{eff}$, of the ${}^{11}$Be nucleus, $4/11$. 

All of these measurements can be addressed within the Halo EFT we will use here. In this theory  the s- and p-wave states of the Beryllium-11 nucleus are generated by core-neutron contact interactions. The theory does not get the interior part of the nuclear wave function correct, but, by construction, it reproduces the correct asymptotics of the wave functions of these states:
\begin{eqnarray}
u_0(r)&=&A_0 \exp(-\gamma_0 r);\nonumber\\ 
u_1(r)&=&A_1 \exp(-\gamma_1 r) \left(1 + \frac{1}{\gamma_1 r} \right),
\label{eq:LOwfs}
\end{eqnarray}
for the $504$ keV and 184 keV states, respectively. As such, it is not a method that is meant to compete with {\it ab initio} calculations of this halo nucleus (see, e.g.~\cite{Fo05,Fo09}) or of ${}^{10}$Be-n scattering~\cite{QN08}. But it could prove complementary to such computations, since Halo EFT provides a way to ensure that the long-distance properties of the halo are correctly taken care of. 

The quantities $\gamma_0$ and $\gamma_1$ are determined by the neutron separation energies of the states in question. At leading order (LO) in the expansion both $A_0$ and $A_1$ are fixed. (In the case of the p-wave this is related to the theorem discussed by Lee at the meeting~\cite{Lee}.) At next-to-leading order (NLO) both $A_0$ and $A_1$ receive corrections whose values must be determined from neutron-$^{10}$Be scattering data. 

\section{Halo EFT for Beryllium-11}

We use the ``Halo EFT" developed in Refs.~\cite{Be02,Bd03} to calculate the properties of the Beryllium-11 nucleus. 
The degrees of freedom in our Halo EFT treatment are the ${}^{10}$Be core and the neutron. The EFT expansion in this case is an expansion in powers of $\omega/B_{\rm high}$. Here $B_{\rm high}$ is, e.g. the excitation energy of states in ${}^{10}$Be, and so is of order a few MeV, and $\omega$ is the energy of the photon exciting the electromagnetic transition of interest. 

\subsection{Strong piece}

In our LO calculation we include the strong s-wave and p-wave interactions that lead to the shallow bound states in the ${}^{11}$Be system through the incorporation of additional spin-zero and spin-one fields:
\begin{eqnarray}
{\cal L}&=&c^\dagger \left(i \partial_t + \frac{\nabla^2}{2 M}\right)c + n^\dagger \left(i \partial_t + \frac{\nabla^2}{2 m}\right)n \nonumber\\
&&+ \sigma^\dagger \left[\eta_0 \left(i \partial_t + \frac{\nabla^2}{2M_{nc}}\right) + \Delta_0\right]\sigma \nonumber \\
&&+ \pi_j^\dagger \left[\eta_1 \left(i \partial_t + \frac{\nabla^2}{2M_{nc}}\right) + \Delta_1\right] \pi_j\nonumber \\
&& - g_0 \left[\sigma n^\dagger c^\dagger + \sigma^\dagger n c\right] \nonumber \\
&&+ \frac{i g_1}{2}\left[\pi_j^\dagger (c \stackrel{\leftrightarrow}{\nabla_j} n)
- (c^\dagger \stackrel{\leftrightarrow}{\nabla_j} n^\dagger) \pi_j\right] + \ldots.
\label{eq:HEFT}
\end{eqnarray}
Here $\ldots$ represents additional p-wave interactions necessary to maintain Gallilean invariance, while
$c$ and $n$ are the ``core" and neutron fields. Hence, $c$ is a bosonic field and $n$ a fermionic one. The field $\sigma$ represents the s-wave state and $\pi_j$ the p-wave state. 

\subsection{Dressing the s-wave state}

\begin{figure}[!h]
\centerline{\includegraphics[width=0.9\columnwidth]{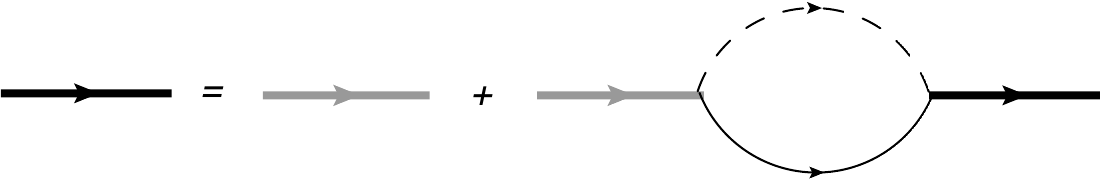}}
\caption{Diagrammatic representation for the Dyson equation which incorporates the one-loop $nc$ dressing of the field representing the s-wave ${}^{10}$Be-n bound state in the theory. Here and below the dashed line represents the field for the ${}^{10}$Be core, and the thin solid line is the neutron. The thick grey line is the undressed $\sigma$ propagator, and the thick black line is the dressed $\sigma$ propagator.}
\label{fig:sigmadressing}       
\end{figure}

In order to treat the shallow s-wave state in the ${}^{10}$Be-neutron system we adopt the counting that has been successfully developed to treat shallow s-wave states in the nucleon-nucleon system~\cite{vK99,Ka98A,Ka98B,Ge98,Bi99}. This can be implemented by noting that then $\sigma n c$ coupling is dimensionful, and taking it to be of order $R_{\rm halo}$.  Meanwhile $n c$ loops will have a typical size of order $1/R_{\rm halo}$~\footnote{In a suitable regularization scheme, e.g. power-law divergence subtraction~\cite{Ka98A,Ka98B}, this is true for both the real and imaginary parts of the loops.}, and so such a counting mandates the resummation of $nc$ loops when computing the $\sigma$ propagator. This can be achieved through the Dyson equation shown in Fig.~\ref{fig:sigmadressing}, which leads to:
\begin{equation}
D_\sigma(p)=
\frac{1}{\Delta_0 + \eta_0[p_0 - {\bf p}^2/(2 M_{nc})] - \Sigma_\sigma(p)},
\label{eq:dressprop}
\end{equation}
with $\Sigma_\sigma(p)$ the one-loop self-energy for the $\sigma$ field. 

This one-loop self-energy is calculated as:
\begin{equation}
\Sigma_\sigma(p)=-\frac{g_0^2 m_R}{2 \pi} \left[\mu + i \sqrt{2 m_R \left(p_0 - \frac{{\bf p}^2}{2 M_{nc}} + i \eta\right)}\right],
\label{eq:Sigsig}
\end{equation}
when computed in power-law divergence subtraction (PDS) with a scale $\mu$~\cite{Ka98A,Ka98B}. Here we have introduced the reduced and total masses of the neutron-core system:
\begin{equation}
M_{nc}=m_n + m_c; \qquad m_R=\frac{m_n m_c}{m_n + m_c}.
\end{equation}

\begin{figure}[htb]
\centerline{\includegraphics[width=0.5\columnwidth]{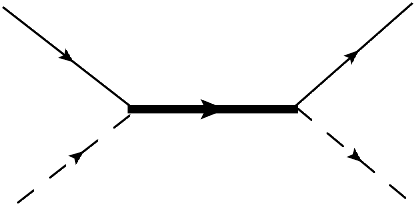}}
\caption{The single Feynman diagram that is needed (once the dressing of the $\sigma$ field has been computed) to obtain the s-wave neutron-core scattering amplitude.}
\label{fig:swavetmatrix}       
\end{figure}

Substituting Eq.~(\ref{eq:Sigsig}) into Eq.~(\ref{eq:dressprop}), we can set the parameters $g_0$ and $\Delta_0$ by computing the s-wave neutron-core scattering amplitude in the theory (\ref{eq:HEFT}) (see Fig.~\ref{fig:swavetmatrix}):
\begin{equation}
t_0(E)=g_0^2 D_\sigma(E,{\bf 0}),
\end{equation}
in the two-body center-of-mass frame. This is then
matched to the effective-range expansion in this channel:
\begin{equation}
t_0(E)=\frac{2 \pi}{m_R}\frac{1}{1/a_0 - \frac{1}{2} r_0 k^2 + i k},
\end{equation}
producing
\begin{equation}
D_\sigma(p)=\frac{2 \pi \gamma_0}{m_R^2 g_0^2} \frac{1}{1 - r_0 \gamma_0} \frac{1}{p_0 - \frac{{\bf p}^2}{2 M_{nc}} + B_0} \mbox{$+$ regular}.
\label{eq:Dsigmadressed}
\end{equation}
In Eq.~(\ref{eq:Dsigmadressed}) the position of the pole is determined by the binding energy $B_0=\gamma_0^2/(2 m_R)$, and $\gamma_0$ is the positive root of the equation:
\begin{equation}
\frac{1}{a_0} + \frac{1}{2} r_0 \gamma_0^2 - \gamma_0=0.
\label{eq:gamma0}
\end{equation}

\subsection{Dressing the p-wave state}

\begin{figure}[!h]
\centerline{\includegraphics[width=0.9\columnwidth]{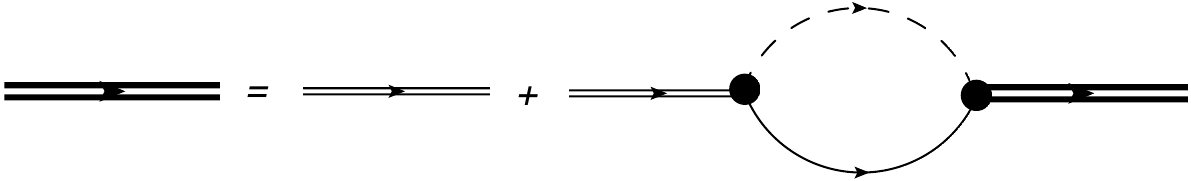}}
\caption{Diagrammatic representation for the Dyson equation which incorporates the one-loop $nc$ dressing of the field representing the p-wave ${}^{10}$Be-n bound state in the theory. Once again, the dashed line represents the field for the ${}^{10}$Be core, and the thin solid line is the neutron. The thin double line is the undressed $\pi$ propagator, and the thick double line is the dressed $\pi$ propagator.}
\label{fig:pidressing}       
\end{figure}

We proceed similarly for the p-wave state. The propagator for this state obeys the Dyson equation depicted in Fig.~\ref{fig:pidressing}. Consequently it takes the form:
\begin{equation}
D_\pi(p)=\frac{1}{\Delta_1 + \eta_1[p_0 - {\bf p}^2/(2 M_{nc})] - \Sigma_\pi(p)}.
\end{equation}
Computation of the one-loop self-energy from the pertinent p-wave vertices in Eq.~(\ref{eq:HEFT}) gives, if PDS is employed:
\begin{equation}
\Sigma_\pi(p)=-\frac{m_R g_1^2 k^2}{6 \pi} \left[\frac{3}{2} \mu + i k\right].
\end{equation}

We note that, since the self-energy loop is cubically divergent, both parameters, $\Delta_1$ and $g_1$, are mandatory for renormalization at LO. This time we are interested in the p-wave core-neutron scattering amplitude in the center-of mass frame:
\begin{eqnarray}
t_1({\bf p}',{\bf p};E)&=&g_1^2 {\bf p}' \cdot {\bf p} D_\pi(E,{\bf 0})\nonumber\\
&=&\frac{6 \pi}{m_R} \frac{ {\bf p}' \cdot {\bf p}}{1/a_1 - \frac{1}{2} r_1 k^2 + i k^3},
\end{eqnarray}
with $k=\sqrt{2 m_R E}=|{\bf p}'|=|{\bf p}|$ for on-shell scattering.
Consequently we obtain:
\begin{equation}
D_\pi(p)=-\frac{3 \pi}{m_R^2 g_1^2}\frac{2}{r_1 + 3 \gamma_1} \frac{i}{p_0 - {\bf p}^2/(2 M_{nc}) + B_1}~\mbox{$+$ regular}.
\end{equation}
Here $\gamma_1=\sqrt{2 m_R B_1}$ is the solution of
\begin{equation}
\frac{1}{a_1} + \frac{1}{2} r_1 \gamma_1^2 + \gamma_1^3=0,
\label{eq:gamma1}
\end{equation}
where $a_1$ is the scattering volume, and $r_1$ the p-wave ``effective range", which, in fact, has dimensions of 1/length.

\subsection{Fixing parameters}

\label{sec-numbers}

Using the experimentally known values $B_0=504$ keV, $B_1=184$ keV, we infer $\gamma_0=0.15$ fm$^{-1}$, and $\gamma_1=0.09$ fm$^{-1}$, which are both of the expected size $1/R_{\rm halo}$. 

There are two possible countings for the p-wave scattering volume $a_1$. Here we adopt that of Ref.~\cite{Bd03}, and take $a_1 \sim R_{\rm halo}^2 R_{\rm core}$, as this requires only one fine tuning in the parameters of the underlying theory. (By comparison, the counting of Ref.~\cite{Be02}, where $a_1 \sim R_{\rm halo}^3$ requires two such fine tunings.) With $a_1 \sim R_{\rm halo}^2 R_{\rm core}$, and $\gamma_1 \sim 1/R_{\rm halo}$,  we can neglect the $\gamma_1^3$ term  in Eq.~(\ref{eq:gamma1}) at leading order and deduce:
\begin{equation}
r_1=-\frac{2}{\gamma_1^2 a_1}.
\label{eq:gamma1lo}
\end{equation}
It follows that if we adopt the central experimental value from Eq.~(\ref{eq:pwavescattlength}) we have $r_1=-0.54$ fm$^{-1}$. (Propagation of experimental errors to the final result is straightforward but will not be discussed in this contribution.) Parametrically $r_1 \sim 1/R_{\rm core}$. At NLO Eq.~(\ref{eq:gamma1lo}) is corrected to:
\begin{equation}
r_1=-\frac{2}{\gamma_1^2 a_1} - 2 \gamma_1,
\end{equation}
which reduces $r_1$ to $-0.72$ fm$^{-1}$, a $\sim$ 30\% correction that is in line with the anticipated expansion parameter of Halo EFT in the ${}^{11}$Be system.

In the s-waves the situation is more straightforward: there we count $a_0 \sim R_{\rm halo}$, and $r_0 \sim R_{\rm core}$. In consequence we can set $r_0=0$ at LO, and obtain from Eq.~(\ref{eq:gamma0}) 
\begin{equation}
\gamma_0=\frac{1}{a_0}.
\end{equation}

Thus, the parameters in the strong sector at LO are $r_1$, $\gamma_0$ (or equivalently $a_0$), and $\gamma_1$. At NLO these are to be supplemented by $r_0$. 

\subsection{Including photons}

Photons are then included in the Lagrangian via minimal substitution:
\begin{equation}
\partial_\mu \rightarrow D_\mu=\partial_\mu + i e Q A_\mu.
\label{eq:minimal}
\end{equation}
The charge operator $Q$ takes different values, depending on whether it is acting on a $c$ field or an $n$ field. $Q\, n=0$ for the neutron, and below we denote the eigenvalue of the operator $Q$ for the $c$ field as $Q\, c=Q_c\, c$. $Q_c=4$ in the case of interest here, where the core is Beryllium-10. 

If magnetic properties are to be discussed we must also consider the insertion of photon interactions with the neutron-core system which are gauge-invariant by themselves (e.g. the contribution of the neutron magnetic moment). But here our focus is on electric properties (and form factors) at LO and NLO, and so it is only necessary to consider how the Lagrangian (\ref{eq:HEFT}) is affected by the substitution (\ref{eq:minimal}). At NNLO in the computation of these properties operators involving the electric field ${\bf E}$ and the fields $c$, $n$, and $\sigma$ which are gauge invariant by themselves contribute to observables. Thus, at that order there is at least one parameter in the Halo EFT description of Coulomb-induced breakup of the Beryllium-11 ground state which cannot be fixed from ${}^{10}$Be-neutron scattering information alone. 

\section{Results for observables}

Using Eq.~(\ref{eq:HEFT}) plus minimal substitution (\ref{eq:minimal}) we obtain a Lagrangian that describes interactions amongst the core, the neutron, the ground and excited states of the ${}^{11}$Be nucleus, and photons. In this section we use this Lagrangian to compute the form factor of the s-wave state and the E1 transition from the s-wave to the p-wave state.
\subsection{s-wave form factor}

\begin{figure}[ht]
\centerline{\includegraphics[width=0.35\columnwidth]{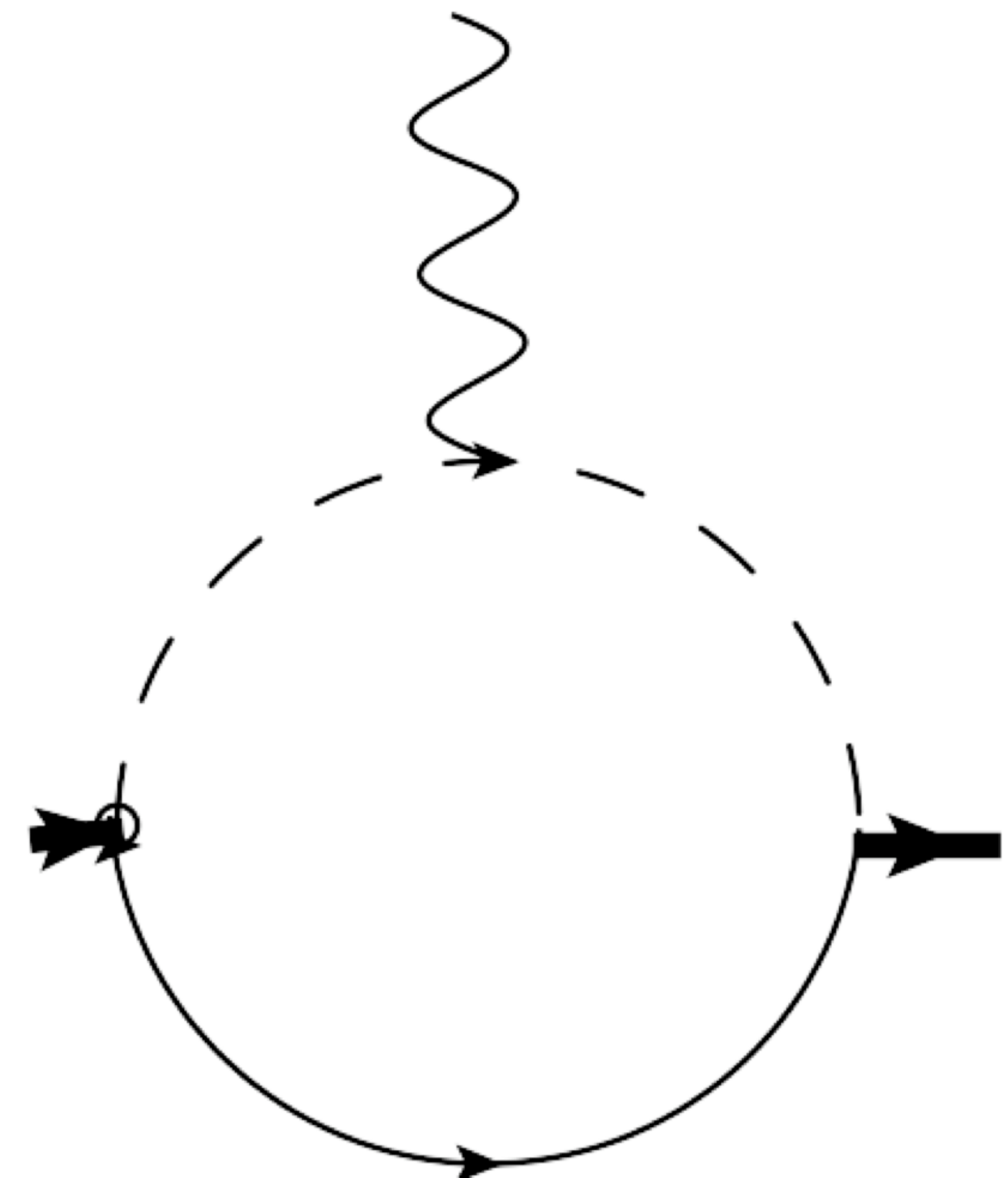}}
\caption{The LO contribution to the irreducible vertex for an $A_0$ photon to couple to the field representing the ${}^{10}$Be-neutron s-wave bound state. Note that there is no diagram for the photon to couple to the neutron as this order, since $Q_n=0$.}
\label{fig:formfactor}       
\end{figure}

The s-wave form factor is computed by calculating the contribution to the irreducible vertex for $A_0 \sigma \sigma$ interactions shown in Fig.~\ref{fig:formfactor}. This is the only diagram it is necessary to consider at leading order. After the application of wave-function renormalization, the irreducible vertex for the $A_0$ photon coupling to the $\sigma$ state is equal to $-i e Q_c G_c(|{\bf q}|)$, where $\bf q$ is the three-momentum of the virtual photon. (Such an interpretation is valid provided the computation is carried out in the Breit frame, where the four-momentum of the virtual photon $q=(0,{\bf q})$.) A straightforward calculation yields: 
\begin{equation}
G_c(|{\bf q}|)= \frac{2 \gamma_0}{f|{\bf q}|} \arctan\left(\frac{f|{\bf q}|}{2 \gamma_0}\right),
\label{eq:Gc}
\end{equation}
with $f=m/M_{nc}=m_R/M$. Note that $G_c(0)=1$, as it should. For the deuteron we have $f=1/2$, and this reduces to the LO result of Ref.~\cite{Ch99}. 

The charge radius of the s-wave state can be extracted according to:
\begin{equation}
G_c(|{\bf q}|)=1 - \frac{1}{6}\langle r_c^2 \rangle {\bf q}^2 + \ldots,
\end{equation}
and an expansion of Eq.~(\ref{eq:Gc}) in powers of $|{\bf q}|$ then yields 
\begin{equation}
\langle r_c^2 \rangle=\frac{f^2}{2 \gamma_0^2}
\end{equation}

Putting in the number for $\gamma_0$ obtained in the previous section we find $\langle r_c^2 \rangle^{1/2}=1.73$ fm from this leading-order HEFT computation. This is about 30\% smaller than the atomic physics measurement of $\langle r_c^2 \rangle^{1/2}=2.463(16)$ fm. (The numbers from the non-energy-weighted sum rule are similar when corrected for $f$.) We expect NLO corrections to increase the charge radius, and a shift of the size needed to produce agreement with experiment is entirely consistent with the nominal $\sim 0.4$ expansion parameter of the Halo EFT in this system.
At NLO a careful treatment of current conservation, which includes an operator associated with gauging the term $\sim \sigma^\dagger \partial_0 \sigma$ in Eq.~(\ref{eq:HEFT}), still yields $G_c(0)=1$, but also produces an increased charge radius, as long as $r_0 > 0$, cf. Ref.~\cite{BS01,Ph02}. The precise size of the increase is fixed once the s-wave effective range $r_0$ is known. 

The charge radius of the p-wave state is also calculable in the Halo EFT. NLO corrections should be smaller there since its binding energy, and so its typical momentum, is lower. To our knowledge there is, as yet, no experimental determination of the charge radius of this state. 

\subsection{E1 transition}

\begin{figure}[!h]
\centerline{\includegraphics[width=0.9\columnwidth]{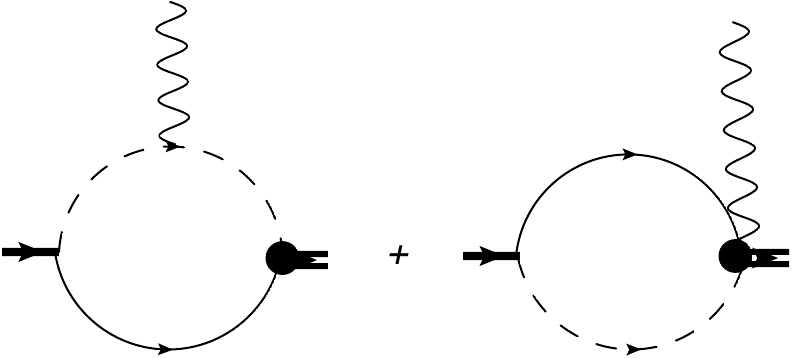}}
\caption{The two diagrams needed for the calculation of the irreducible vertex that governs the s-to-p-state transition, $\Gamma_{j \mu}$ in Halo EFT at leading order.}
\label{fig:Gammajmu}       
\end{figure}

Now we consider the E1 transition from the s-wave state to the p-wave state. The irreducible vertex for this transition is depicted in Fig.~\ref{fig:Gammajmu}. Here we compute the transition for a photon of arbitrary four momentum $k=(\omega,{\bf k})$, and the sum of diagrams yields $-i \Gamma_{j \mu}$ where $j$ is the polarization index of the p-wave state and $\mu$ that of the photon.

We first observe that both the diagrams depicted in Fig.~\ref{fig:Gammajmu} are divergent, but that the divergences cancel, as they should since gauge invariance precludes us from writing down any contact interaction that contributes to this observable at leading order. Current conservation at LO can be explicitly checked, and we find that, as long as both diagrams are considered:
\begin{equation}
k^\mu \Gamma_{j \mu}=0.
\label{eq:currentcons}
\end{equation}
Note that if only the long-distance E1 mechanism on the left-hand side of Fig.~\ref{fig:Gammajmu} is considered, as was done, for example, in Ref.~\cite{TB08}, then current conservation is not satisfied, and it appears that some input from short-distance physics is needed in order to define the prediction for this observable. 

For real photons we have ${\bf k} \cdot {\bf \epsilon}=0$ and ${\bf k} \cdot {\bf q}=0$, with ${\bf q}$ the incoming momentum of the s-wave state. With these restrictions we can write the space-space components of the vertex function as:
\begin{equation}
\Gamma_{ji}=\delta_{ji} \Gamma_E + k_j q_i \Gamma_M.
\end{equation}
Current conservation (\ref{eq:currentcons}) then provides an alternative way to calculate $\Gamma_E$, it tells us that:
\begin{equation}
\omega \Gamma_{j0}={\bf k}_j \Gamma_E.
\end{equation}

But, for $\Gamma_{j0}$, the diagram on the right of Fig.~\ref{fig:Gammajmu} need not be considered, and so
\begin{equation}
\Gamma_{j0}({\bf k}) \sim  \int d^3r \frac{u_1(r)}{r} Y_{1j}(\hat{r}) e^{i {\bf k} \cdot {\bf r}} \frac{u_0(r)}{r},
\label{eq:Gamma0j}
\end{equation}
where $u_1$ and $u_0$ are the leading-order wave functions of the s- and p-wave states, given by Eq.~(\ref{eq:LOwfs}).
As $|{\bf k}| \rightarrow 0$ Eq.~(\ref{eq:Gamma0j}) reduces to:
\begin{equation}
\Gamma_{j0}({\bf k}) \sim {\bf k}_j \int dr u_1(r) r u_0(r),
\end{equation}
an equation in which, of course, the integral is the canonical form of the E1 matrix element. 

Using the definition of B(E1)~\cite{TB05} we find that the B(E1) strength for this transition is related to the renormalized, irreducible vertex $\bar{\Gamma}_E$ by:
\begin{equation}
B(E1)=\frac{1}{4 \pi} \left(\frac{\bar{\Gamma}_E}{\omega}\right),
\end{equation}
where 
\begin{eqnarray}
\bar{\Gamma}_E &\equiv& \sqrt{Z_0 Z_1} \Gamma_E\\
&=&\frac{1}{m_R^2 g_0 g_1} \sqrt{\frac{-12 \pi^2 \gamma_0}{r_1}}  \Gamma_E
\end{eqnarray}
at leading order. 

Hence, 
\begin{equation}
B(E1)=\frac{Z_{eff}^2 e^2}{4 \pi} \frac{4 \gamma_0}{-3 r_1} \left[\frac{2 \gamma_1 + \gamma_0}{(\gamma_0 + \gamma_1)^2}\right]^2
\label{eq:BE1}
\end{equation}
is the leading-order Halo EFT result. 
No cutoff parameter is needed in order to get a finite result for B(E1): our value is finite without regularization. We note that the result (\ref{eq:BE1}) is ``universal" in the sense that it applies to any E1 s-to-p-wave transition in a one-neutron halo nucleus. Once $r_1$, $\gamma_1$, and $\gamma_0$ are known for a given one-neutron halo the prediction (\ref{eq:BE1}) is accurate up to corrections of order $R_{\rm core}/R_{\rm halo}$.

In the case of ${}^{11}$Be we insert the numerical values from Sec.~\ref{sec-numbers} and find:
\begin{equation}
{\rm B}_{\rm LO}{\rm (E1)}=0.085~e^2~{\rm fm}^2.
\end{equation}
This is about 20\% too low compared to the value (\ref{eq:CEXBE1}) obtained in Ref.~\cite{Su07}.

NLO corrections from the wave-function renormalization factors associated with the s-wave and p-wave fields will serve to increase the leading-order prediction. We expect them to be of the size necessary to bring the prediction into agreement with the datum (\ref{eq:CEXBE1}). As was discussed above in the case of $r_c$, there is a prediction for B(E1) at NLO if we can fix the value of $r_0$ from some other data. In particular, we are optimistic that we can fix the magnitude of $r_0$ by examining experimental results for the low-energy E1 strength function in breakup to the ${}^{10}$Be-neutron channel: $d$B(E1)$/dE$. The computation of that observable is a straightforward extension of the s-wave bound state to p-wave bond state E1 calculation carried out here. 

We emphasize that if the s-wave effective range, $r_0$, is known then the NLO result is predictive. The first contribution of physics at scale $R_{\rm core}$ that cannot be fixed from hadronic-interaction observables does not enter the result for B(E1) until next-to-next-to-leading order, i.e. it is suppressed by $B_{\rm lo}/B_{\rm hi} \sim (R_{\rm core}/R_{\rm halo})^2$. 

\section{Conclusion}

This brief discussion of electromagnetic observables in the Beryllium-11 system already displays the significant recent experimental activity that has been focused on this nucleus. Coulomb excitation has been used at a variety of facilities to probe E1 transitions, and atomic-physics experiments have made great strides through advances in trapping technology. This means that the time is ripe for a detailed analysis of electromagnetic properties of halo systems. Here we have shown how EFT can provide such an analysis. 

The LO result in the EFT mirrors elegant analytic approaches to halo nuclei (see, e.g. Refs.~\cite{TB04,TB08,TB05}). But, in contrast to those works, there is no regulator dependence in the LO result for the E1 strength. This is a consequence of current conservation in our formalism. Moreover, the correction of $O(R_{\rm core}/R_{\rm halo})$ to the LO prediction can be systematically calculated using EFT techniques.  We note that there are no spectroscopic factors in our approach. They represent the effect of physics at scale $R_{\rm core}$, and the order at which such ``short-distance physics" impacts electromagnetic observables can be delineated in the EFT. In the Beryllium-11 system the EFT expansion parameter is $\sim 0.4$, and the LO predictions for $r_c$ and B(E1) obtained here are in agreement with experimental data to that expected level of accuracy. Improving this LO prediction to NLO will require us to fix $r_0$, the effective range for s-wave neutron-core scattering, from experiment. This can perhaps be done using data on dissociation of the Beryllium-11 nucleus to the continuum.

The Halo EFT therefore provides a complement to {\it ab initio} methods~\cite{Fo05,Fo09,QN08}, which can struggle to describe E1 transitions and radii in these extended systems because of the widely varying core and halo scales that are present in the problem. In Halo EFT this wide separation of scales is the basis for the calculation. Input that summarizes the physics at scale $R_{\rm core}$ can be taken from either simulation or experiment, and the EFT is then used to predict the outcome of experiments that probe dynamics at the halo scale. The application of this approach to other one-neutron halos, and to two-neutron halos such as ${}^{11}$Li, are obvious next steps.

\section*{Acknowledgments}
This research was supported by the US Department of Energy under grant DE-FG02-93ER40756, by the BMBF under contracts
06BN411 and 06BN9006, and by the Mercator programme of the Deutsche Forschungsgemeinschaft. 
DRP thanks the organizers of the conference for a stimulating meeting, and for a very enjoyable time in Bonn. 


\end{document}